\documentclass[aps,prl,twocolumn,floatfix, showpacs]{revtex4}
\usepackage{graphicx}
\usepackage{dcolumn}
\usepackage{bm}

\newcommand{\be}{\begin{equation}}
\newcommand{\ee}{\end{equation}}

\begin{document}

\title{Enhanced azimuthal rotation of the large-scale flow through stochastic cessations in turbulent rotating convection with large Rossby numbers}
\author{Jin-Qiang Zhong}
\author{Hui-Min Li}
\author{Xue-Ying Wang}
\affiliation{School of Physics Science and Engineering, Tongji University, Shanghai, China}
\date{\today}
 
\begin{abstract}
We present measurements of the azimuthal orientation $\theta(t)$ and thermal amplitude $\delta(t)$ of the large-scale circulation (LSC) of turbulent rotating convection within an unprecedented large Rossby number range $1{\le}$Ro${\le}314$. Results of $\theta(t)$ reveal persistent rotation of the LSC flow in the retrograde direction over the entire  Ro range. The rotation speed ratio remains a constant of $0.13\pm0.01$ for $10{\le}$Ro${\le}70$, but starts to increase with increasing Ro when Ro$>$70. We identify the mechanism through which the mean retrograde rotation speed can be enhanced by stochastic cessations in the presence of weak Coriolis force, and show that a low-dimensional, stochastic model provides predictions of the observed large-scale flow dynamics and interprets its retrograde rotation.          
\end{abstract}

\pacs{47.27.te, 05.65.+b, 47.27.eb}

\maketitle

Turbulent convection in the presence of rotation occurs in a variety of natural flows in the atmosphere and the oceans \cite{Va06}. Large-scale coherent flows often exist in these geophysical fluid systems and play crucial roles in their turbulent heat and mass transport. Examples include the thermohaline circulation in the oceans \cite{MS99}, and the tropical atmospheric circulation (the Hadley cell) \cite{ENB94}.  The canonical framework to study turbulent convection flow is the Rayleigh-Benard convection (RBC) system, i.e., a horizontal fluid layer heated from below and cooled from above \cite{AGL09, LX10}. In turbulent RBC, thermal plumes that emanate from the thermal boundary layers (BL) are organized spontaneously into a large-scale circulation (LSC) (see e.g. \cite{KH81, NBS01, BNA05, XZZCX09}). Laboratory experiments reveal that the LSC survives under the influence of modest rotations if the ratio of the buoyancy and Coriolis forces, expressed in the Rossby number (Ro), is greater than one \cite{KCG08, ZA10}. 

A prominent dynamical feature of the LSC in turbulent rotating RBC in cylindrical samples is the azimuthal rotation of its polar circulation plane with nearly a constant velocity \cite{HKO02, BNA05, BA06, SXX05, KCG08, ZA10}. Developed from the fluid momentum equations, one-dimensional theoretical models suggested that when the LSC flow rotates steadily in the azimuthal direction, the Coriolis force, which accelerates the LSC rotation, is balanced by the viscous drag from the kinetic BLs \cite{BA06, KCG08}. These models predicted a constant retrograde rotation speed ${\omega}$ close to the sample rotating rate $\Omega$ (dashed line in Fig. 1). The model prediction was shown to be in reasonable consistency with the experimental measurements of ${\omega}$ (open circle, Fig. 1) in the presence of the Earth's Coriolis force ($\Omega$=$\Omega_E$=7.3${\times}10^{-5}$rad/s, \cite{BA06}).  Experiments \cite{HKO02, KCG08, ZA10, ZSL15} with deliberate rotations (but with $\Omega{\ge}$100$\Omega_E$) reported LSC rotation speeds $\omega$ about one order in magnitude less than $\Omega$ (solid symbols, Fig. 1). Moreover, conflicting results of $\omega{>}\Omega$ had been reported in other experiments with $\Omega$=$\Omega_E$ \cite{SXX05, XZX06}. Although controversial results of the LSC rotation were obtained, these experimental studies revealed, from different perspectives, a variety of azimuthal dynamics of the LSC (such as cessations and erratic rotations) that has not been accounted for in the aforementioned one-dimensional model. Our understanding in the fundamental dynamics of the large-scale flow in rotating turbulent convection is far from complete.

\begin{figure}
\includegraphics[width=3.0in]{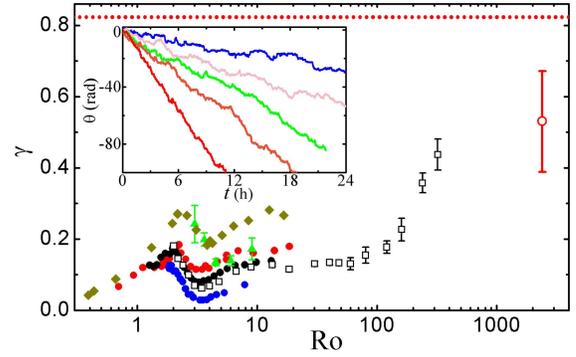}
\caption{Results for the LSC rotation speed ratio ${\gamma}$=$\omega/{\Omega}$ as a function of Ro. Solid circles: data from \cite{ZA10} with Pr=4.38 and Ra=1.8${\times}10^{10}$(red), 8.97${\times}10^9$(black) and 2.25${\times}10^9$(blue); Diamonds: data from \cite{HKO02}; Triangles: data from \cite{KCG08}; Open circle: data from \cite{BA06} with $\Omega$=${\Omega}_E$; Squares: this work. The error bars are the rms spread of $\theta(t)$ around the linear fit. Dashed line: theoretical prediction of $\gamma$=1/(1+12Re$^{-1/2}$) from \cite{BA06} with Pr=4.38 and Ra=8.24${\times}10^{9}$. Inset: Examples of $\theta(t)$ as functions of time for Ro=13.1(red), 18.2(orange), 39.4(green), 59.0(magenta) and 236(blue).}
\label{fig1:}
\end{figure}

\begin{figure}
\includegraphics[width=2.8in]{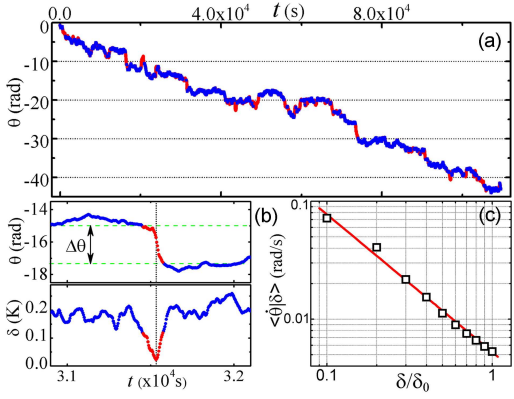}
\caption{(a) A long time series of $\theta(t)$ for Ro=236. (b) Time series of both $\theta(t)$ and $\delta(t)$ during cessations. Sections in the time series with $\delta$ larger (smaller) than $\delta_c=0.5\delta_0$ are shown in blue (red) color. The vertical dashed line in (b) indicate the moment when $\delta$ is minimum. (c) Conditional mean $\langle{\dot{\theta}}|{\delta}\rangle$ as a function of $\delta/\delta_0$. The straight line indicates $\langle{\dot{\theta}}|{\delta}\rangle$=$\dot{\theta}_0(\delta/\delta_0)^{\alpha}$ with $\dot{\theta}_0$=5.3$\times10^{-3}$rad/s and $\alpha=-1.16$.} 
\label{fig2:}
\end{figure}

In this paper we present measurements of the azimuthal orientation $\theta(t)$ of the large-scale flow in an unprecedented large Ro range $1{\le}$Ro${\le}314$. In the low-Ro range, results of the LSC rotation speed ratio $\gamma$ agree well with existing experimental data. In the large-Ro range, we observe that $\dot{\theta}(t)$ can be largely enhanced due to intense accelerations during small-amplitude (cessation) events, and report a heretofore unanticipated increasing of $\gamma$ when Ro$\ge$70. We elucidate that the observed phenomena of retrograde rotations of the large-scale flow can be understood using a low-dimensional stochastic model.  

The experiment was performed using an apparatus described before \cite{ZSL15}. We used a cylindrical cell with a height $L$=24.0cm and an aspect ratio of 1.00. The sample was filled with deionized water at a mean temperature of $40.00^{\circ}$C. The Rayleigh number, Ra${\equiv}{\alpha}g{\Delta}TL^3/{\kappa}{\nu}$=8.24${\times}10^9$ and the Prandtl number Pr${\equiv}{\nu}/{\kappa}$=4.38 remained constant. Here $g$ is the gravitational acceleration,  ${\alpha}, {\nu}$ and ${\kappa}$ are the thermal expansion coefficient, the viscosity and thermal diffusivity respectively, and  ${\Delta}T$ is the applied temperature difference. The sample had a 4mm thick, Plexiglas sidewall. Eight thermistors, equally spaced azimuthally in the horizontal midplane of the cell, were placed into the sidewall. We measured the temperature of each thermistor $T_i$, and fitted them with the function $T_i$=$T_0$+${\delta}$cos$(i{\pi}/4-\theta), i$=1,...8. Following this experimental protocol \cite{BNA05}, we determined the LSC thermal amplitude $\delta$ and the azimuthal orientation $\theta$ of its circulating plane. The convection cell was mounted on a rotary table with high-quality performance for slow-rotating experiments (see Supplemental Material for detailes \cite{SM}). The rotating velocity of the table $\Omega$ can be selected within (7.8${\times}10^{-4}$,0.1) rad/s, so Ro${\equiv}\sqrt{{\alpha}g{\Delta}T/L}/2{\Omega}$, covered the range $1{\lesssim}$Ro${\le}314$. 

Over the entire Ro range studied, retrograde rotation of the LSC plane was observed that led to linear decrease of $\theta$ in time. Figure 1 shows the time series of $\theta(t)$ for several Ro. When Ro increases the mean rotating rate $\omega$=${\vert}{\langle}\dot{\theta}{\rangle}{\vert}$, determined through linear fits to the traces of $\theta(t)$, decreases monotonically.   
The ratio of  the LSC rotation speed, ${\gamma}$=$\omega/{\Omega}$ is depicted  as a function of Ro. 
We first observed complicated Ro-dependence of ${\gamma}$ in the range $1{\le}$Ro${\le}10$ that agree with previous works \cite{HKO02, ZA10}. In the range of $10{\le}$Ro${\le}70$, $\omega$ appears to be proportional to ${\Omega}$, yielding ${\gamma}$=0.13${\pm}$0.01 independent of Ro. When Ro${\ge}70$, ${\gamma}$ gradually increases and reaches a value of $0.42$ for Ro=314. This Ro corresponds to the slowest sample rotating rate $\Omega{\approx}10{\Omega}_E$ we applied.  Result from \cite{BA06} with $\Omega$=${\Omega}_E$ is shown by the rightmost circle.


\begin{figure}
\includegraphics[width=3.0in]{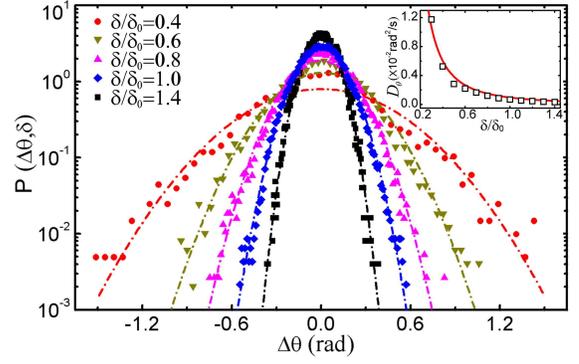}
\caption{The PDFs of the azimuthal displacement $\Delta{\theta}$ for given amplitudes ${\delta}$ with Ro=236. Dashed lines: Gaussian fit to data. Inset: Diffusivity $D_{\theta}$ as a function of ${\delta}/{\delta_0}$. The solid curve is a power-function fit: $D_{\theta}$=$D_0{\delta}^{-2}$ with $D_0$=5.2${\times}10^{-5}$rad$^2$K$^2$/s.
}
\label{fig3 :}
\end{figure}  

Figure 2 shows a long time series of $\theta(t)$ for Ro=236.  A mean retrograde rotation speed of $\omega$=$0.36\Omega$ is observed in this example. We find that a broad spectrum of the LSC azimuthal dynamics contributes to the net rotation, and determines $\omega$. Over large time scales of hours $\theta(t)$ appears to rotate steadily at a nearly constant velocity. In a short time scale of minutes the LSC orientation undergoes meandering motion that resembles Brownian motions. Moreover, there are occasional violent accelerations in $\theta(t)$ (shown in red color) that lead to large instantaneous azimuthal velocities of up to 600 times ${\langle}\dot{\theta}{\rangle}$. These fast reorientation events occur intermittently, accompanied by decreasing of the LSC amplitude $\delta$ that typify cessation events \cite{BNA05} (Fig. 2b). The relationship between the instantaneous azimuthal velocities $\dot{\theta}$ and $\delta$ can be quantitatively presented by the conditional expectation $\langle{\dot{\theta}|\delta}\rangle$, i.e., the time-average of $\dot{\theta}$ for a given value of $\delta$. As depicted in Fig. 2c, $\langle{\dot{\theta}|\delta}\rangle$ increases when  $\delta$ decreases and can be best represented as $\langle{\dot{\theta}}|{\delta}\rangle$=$\dot{\theta}_0(\delta/\delta_0)^{-1.16}$.        

The enhanced azimuthal rotation velocity $\dot{\theta}$ when $\delta$ is small implies that the LSC rotational inertia in its circulating plane plays an important role to determine its azimuthal rotation. To interpret this phenomenon, we develop a Langevin equation from the Navier-Stokes equation to describe the motion of  $\theta(t)$ in the presence of rotations \cite{BA07, ZSL15, AAG12}:
\be
\ddot{\theta}=-\frac{\delta}{\tau_{\dot{\theta}}\delta_0}\dot{\theta}+\frac{\delta{\Omega}}{\tau_{\dot{\theta}}\delta_0}+f_{\dot{\theta}}(t)
\ee
Here the azimuthal fluid acceleration, expressed in ${\ddot{\theta}}$, is given by its rotational inertia with a damping time scale $\tau_{\dot{\theta}}$=$L^2/2{\nu}Re$ and the Coriolis force. $\delta_0$=${\langle}\delta{\rangle}$ is the time-average amplitude. The background turbulent fluctuation is modelled by a delta-correlated, Gaussian noise term $f_{\dot{\theta}}$ that has zero mean and an intensity $\Gamma_{\theta}$=$D_\theta/\tau_{\dot{\theta}}^2$. 

Equation 1 suggests that when $\delta$ decreases, the inertial damping, ${\delta}{\dot{\theta}}/{\delta_0}{\tau_{\dot{\theta}}}$, becomes smaller so the LSC plane is accelerated by the turbulent fluctuations and moves more freely in the azimuthal direction. Here we examine the angular displacement $\Delta\theta(dt)$=$\tilde{\theta}(t+dt)$-$\tilde{\theta}(t)$ of the detrended LSC orientation $\tilde{\theta}(t)$=${\theta}(t)$-$\omega{t}$, and determine the conditional probability distribution $P(\Delta\theta,\delta)$ using time series $\theta(t)$ with $\delta$ restricted to the range ($\delta$-0.1$\delta_0$, $\delta$+0.1$\delta_0$). Figure 3 shows results of $P(\Delta\theta,\delta)$ for several $\delta$ with a time interval $dt$=40s. All of the distributions have a Gaussian shape with an increasing variance when $\delta$ decreases. They indicate that azimuthal motion of the LSC plane is well characterized by Brownian motion irrespective of the LSC strength $\delta$. 

Since the LSC amplitude $\delta$ varies relatively slowly with a characteristic time scale $\tau_{\delta}$ much larger than $\tau_{\dot{\theta}}$ \cite{BA08a}, $P(\Delta\theta,\delta)$ reveals statistically how diffusive motion of $\theta(t)$ depends on $\delta$. Theoretically  $P(\Delta\theta,\delta)$ can be determined using the steady-state solution of the Fokker-Planck Equation corresponding to Equ. (1) \cite{AAG11}: $P(\Delta\theta,\delta)$=$[2\pi{\sigma_{\theta}}^2(dt)]^{-1/2}$exp$[-(\Delta\theta$-${\langle}\Delta\theta{\rangle})^2$/$2{\sigma_{\theta}}^2(dt)]$. With a significantly long sampling time interval $dt{\gg}\tau_{\dot{\theta}}{\delta_0}/\delta$, its variance is given by ${\sigma_{\theta}}(dt)$=$\sqrt{D_{\theta}{\cdot}dt}$, where $D_{\theta}$=$\Gamma_{\theta}\tau^2_{\dot{\theta}}{\delta^2_0}/\delta^2$ is the diffusivity of $\theta$ \cite{SM}.  Experimentally we measured ${\sigma_{\theta}}^2(dt)$=${\langle}(\Delta\theta$-${\langle}\Delta\theta{\rangle})^2{\rangle}$ with $dt$ that spans a large range $(0.5\tau{\le}dt{\le}5\tau)$, and determine $D_{\theta}$ as a function of $\delta$. As shown in the inset to Fig. 3, $D_{\theta}(\delta)$ increases rapidly with decreasing $\delta$ and follows closely a simple power function: $D_{\theta}(\delta)$=$D_0{\delta}^{-2}$. Such a power law dictates $D_{\theta}({\delta})$ for all Ro in the range Ro${\ge}5$. The close agreement between the theory and experimental data indicates that the diffusivity of $\theta(t)$ can be significantly enhanced when during small-amplitude events.

To better understand the small-amplitude behavior of the LSC, we measured the occurrence frequency $f_c$ of cessations, which was defined to occur when ${\delta}{<}\delta_c$=$0.5{\delta_0}$ \cite{commentCESS}. 
In order to ensure sufficient statistics of $f_c$ we recorded a long time series ($\sim$100 hours) for each Ro in the high-Ro range. Results for $f_c$ is depicted in Fig. 4c. One sees that $f_c$ remains roughly constant in the large-Ro range (Ro$\gtrsim$5). In the fast rotating range when Ro$\lesssim$5, $f_c$ increases sharply with increasesing $\Omega$. 

The dynamics of the LSC amplitude ${\delta}(t)$ is governed by a second Langevin equation \cite{BA07, ZSL15, AAG12}:
\be
 \dot{\delta}=\frac{BD_{\delta}}{2\delta_0}+\frac{\delta}{\tau_\delta}-\frac{\delta^{3/2}}{\tau_{\delta}\delta_0^{1/2}}+f_{\delta}(t)
 \ee
This amplitude equation, combined with Eq. (1), constitutes the full LSC model.  Here the increment of the LSC strength $\dot{\delta}$ is given by the buoyancy force, the thermal diffusion and the viscous dissipation from the BLs. The constant $B$ reveals the change rate $\dot{\delta}$ when ${\delta}\ll{\delta_0}$. $f_{\delta}(t)$ presents Gaussian noise with intensity $\Gamma_{\delta}$=$D_{\delta}/{\tau^2_{\delta}}$. 

\begin{figure}
\includegraphics[width=3.2in]{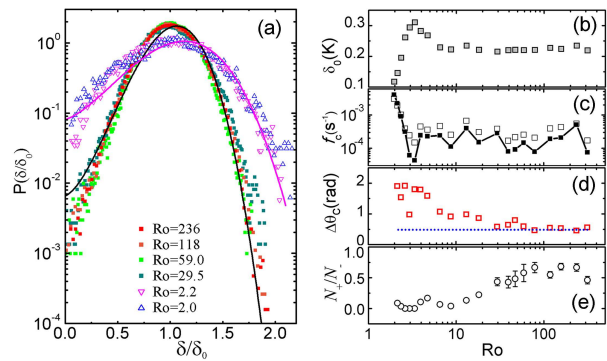}
\caption{Some dynamical properties of the LSC during cessations. (a) The PDFs of $\delta/\delta_0$. The two solid curves are theoretical fittings of $P(\delta/\delta_0)$ for Ro=236 (black) and Ro=2.2 (magenta). (b) Mean amplitude $\delta_0$ as a function of Ro.  (c) $f_c$ as a function of Ro. Open squares: experimental data; Solid squares: theoretical results. (d) Azimuthal displacement ${\Delta}\theta_c$. The dashed line shows the theoretical estimate of ${\Delta}\theta_d$ due to diffusion for Ro=236. (e) Ratio of $N_+/N_-$. Error bars: the standard deviations. 
}
\label{fig4 :}
\end{figure} 

The stochastic behavior of $\delta(t)$ is described by diffusive motions in a potential well \cite{Ga04}:  $V(\delta)$=$-BD_{\delta}{\delta}/2\delta_0$-${\delta}^2/2{\tau_\delta}$+$2{\delta}^{5/2}/5{\tau_\delta}\delta_0^{1/2}$.  The PDF of $\delta/{\delta_0}$ is: $P(\delta/\delta_0)$
=$(\sqrt{2{\pi}}\sigma_{\delta})^{-1}$exp$[-2V(\delta/\delta_0)$/$D_\delta]$ with a variance $\sigma_{\delta}$=$\sqrt{D_{\delta}\tau_{\delta}}/\delta_0$ that depends on $\delta_0$. Results of $P(\delta/\delta_0)$ are presented in Fig. 4a for several Ro. When Ro${\gtrsim}$5 data of $P(\delta/\delta_0)$ collapse into one single curve. As shown in Fig. 4b, in this range the Coriolis force is so weak that it has not yet altered the LSC strength $\delta_0$, thus the variance of $P(\delta/\delta_0)$ remains constant. When Ro${\lesssim}$5, the half width of the PDF starts to increase with decreasing Ro. In this range $\delta_0$ decreases rapidly, indicating the decay of the LSC strength in the background of strong rotations \cite{KCG08, ZA10}. We determine the parameters ($B, D_{\delta}, \tau_{\delta}$) in Equ. (2) as functions of Ro through theoretical fittings of $P(\delta/\delta_0)$ to the experimental data, and compute theoretically the cessation frequency $f_c$, using the Arrhenius formula \cite{SM}. Theoretical  predictions of $f_c$ are compared with the experimental data in Fig. 4c. They imply that  that the observed Ro-dependence of $f_c$ can be well predicted by Equ. (2).


Also of interest is the mean angular change $\Delta\theta_c$ of the LSC during cessations (Fig. 2b). Theoretically we estimate the displacement $\Delta\theta_d$ due to diffusion of $\theta(t)$ within a mean cessation duration $\Delta{\tau}$, based on the diffusivity $D_{\theta}(\delta)$ shown in Fig. 3 \cite{SM}. Results of $\Delta\theta_d$ for Ro=236 is shown by the dashed line in Fig. 4d and compared with the experimental data of $\Delta\theta_c$. We see that when Ro$\ge$70, the angular change during cessations are mainly determined by diffusive motion $\Delta\theta_c{\approx}\Delta\theta_d$. However,  when Ro${\le}70$, $\Delta\theta_c$ becomes larger than $\Delta\theta_d$ because the mean drift velocity of ${\langle}\dot{\theta}{\rangle}$ due to the Coriolis force starts to exceed $\Delta\theta_d/{\Delta}\tau$.   

The angular displacement $\Delta\theta_c$ has a preference in the retrograde direction. It is readily seen in Fig. 2a that retrograde rotations during cessations are more frequent than the prograde ones. 
For a quantitative analysis we count the number of the cessation events, $N_+(N_-)$, if the net displacement $\Delta\theta_c$ is in the prograde (retrograde) direction. The ratio of $N_+/N_-$ is depicted in Fig. 4e as a function of Ro. We see that $N_+/N_-{\approx}0.6{\pm}0.1$ when Ro${\ge}70$, but decreases with decreasing Ro. The underlying mechanism for the observed directional azimuthal displacements during cessations is not well understood. We infer that it is associated with the gradual variation in $\delta$ when cessations occur. As predicted by Equ. (2), during cessations the change rate, $|\dot{\delta}|$=$dV/d\delta|_{\delta{\approx}0}$=$BD_{\delta}/2{\delta_0}$, remains finite and independent of ${\delta}$ \cite{BA08a, BA06b}. The traces of ${\delta}(t)$ exhibit a wedge shape as shown in Fig. 2b. Thus the LSC retains in substantial strength. The fluid momentum produced by the Coriolis force, ${\delta}{\Omega}/\tau_{\dot{\theta}}\delta_0$, continues to accelerate the LSC rotations preferentially in the retrograde direction.    
 
Our interpretation for the observed increasing rotation speed ratio $\gamma$ in a large-Ro range is as follows: When small-amplitude events (such as cessations) occur the inertial damping for the LSC azimuthal motion decreases. The diffusivity of ${\theta}$, $D_{\theta}$=$D_0{\delta}^{-2}$, increases accordingly and gives rise to large azimuthal velocities $\dot{\theta}$. The dynamical properties of cessations, including the occurrence frequency $f_c$ and the angular displacement $\Delta{\theta}_c$ are, however, independent of Ro when Ro$\ge5$, since the Coriolis force has not yet altered the LSC strength $\delta$ under weak rotations (Fig. 4b-d). Thus with increasing Ro, the angular displacements during cessations take a larger part in the LSC azimuthal motion and increase ${\langle}\dot{\theta}{\rangle}$ further. An increasing in $\gamma$ should occur when the enhanced ${\langle}\dot{\theta}{\rangle}$ due to cessations, exceeds the mean drift velocity ascribed to the Coriolis force along. The cessation-enhanced $\gamma$ can be estimated based on the measured properties:  ${\Delta}{\gamma_c}$=$f_c{\Delta}{\theta_c}(N_-$-$N_+)/\Omega$. Results for ${\Delta}{\gamma_c}$ are shown in Fig. 5a. We see that ${\Delta}{\gamma_c}$ is insignificant with small Ro, but becomes important when Ro${\ge}$70.   

To clarify how the mean azimuthal rotation speed is dependent on $\delta$ for various Ro, we show in Fig. 5b the conditional mean $\langle{\gamma}|{\delta_l}\rangle$, determined by time series of $\theta(t)$ with all small-amplitude events ($\delta{<}\delta_l$) removed \cite{SM}. We find that $\langle{\gamma}|{\delta_l}\rangle$ decreases with increasing $\delta_l$. The reduction in $\gamma$ is greater with a larger Ro. The different curves of $\langle{\gamma}|{\delta_l}\rangle$ (Ro) appear to converge to a low value of $0.06{\pm}0.02$ when $\delta_l{\approx}\delta_0$. To account for the observed slow retrograde rotation velocity, we suggest that an additional viscous force, $f_{K}$=$-K(\delta)\dot{\theta}/{\tau}_{\dot{\theta}}$, is involved in Eq. (1). Such a greater level of viscous damping has been attributed to turbulent viscosity \cite{HKO02, ZA10}. We propose here that $K(\delta)$ is sensitive to the LSC strength $\delta$. When $\delta{\ll}\delta_0$, $K(\delta){\ll}1$, the diffusivity of $\theta$ is mainly given by the inertial damping and follows $D_{\theta}(\delta)$=$D_0{\delta}^{-2}$. When $\delta{\ge}\delta_0$, $K(\delta){\gg}1$ and $f_{K}$ dominates other dissipation forces. The LSC rotation speed ratio $\gamma{\approx}1/K$  becomes one order in magnitude less than one, and is nearly independent of Ro. Finally we find that the reduction of $\gamma$ when removing the cessation events, ${\Delta}{\gamma}$=$\langle{\gamma}|0\rangle$-$\langle{\gamma}|{\delta_c}\rangle$, agrees with our estimate of ${\Delta}{\gamma_c}$ as shown in Fig. 5a. 

 \begin{figure}
\includegraphics[width=3.4in]{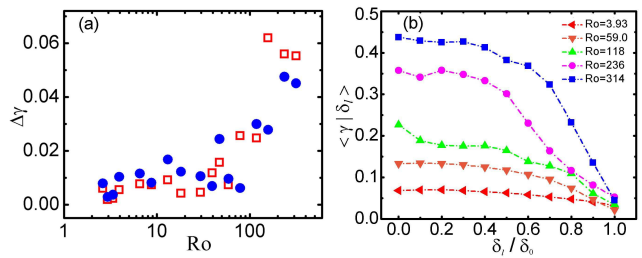}
\caption{(a) The enhanced ratio $\Delta{\gamma}$ due to cessations. Open squares: experimental data; Solid circles: theoretical estimate of ${\Delta}\gamma_c$. (b) The rotation speed ratio $\gamma$ as functions of $\delta_l/\delta_0$. Results obtained using time series of $\theta(t)$ with $\delta{\ge}\delta_l$.}
\label{fig5 :}
\end{figure}  

In summary, we report persistent retrograde rotation of the LSC flow in turbulent RBC within a large Ro number range 1$\lesssim$Ro$\le$314. The rotation speed ratio $\gamma$ is observed to increases with increasing Ro when Ro$>$70. With close inspections of the stochastic behavior of the flow we elucidate that during small-amplitude events (cessations) the azimuthally rotating velocity of the flow can be significantly enhanced. We remark that the dynamics of azimuthal rotation of large-scale flows driven by Coriolis force, besides of its ubiquity in fluid flows in the atmosphere and the oceans, is one of the important ingredients in self-excited dynamo theories \cite{RG00}. For example, it is suggested that in the Earth's fluid core a meridional circulation created by turbulent thermal convection within the inner core tangent cylinder is deflected by the Earth's Coriolis force and turns into an azimuthally rotating flow near the inner-core boundary (ICB) \cite{GR95, OCG99}. The azimuthal strength of this large-scale flow plays a crucial role in geomagnetic field generation since it effectively shears the existing poloidal field and generates new toroidal field in the fluid core near the ICB. Moreover, the resultant azimuthal flows produce magnetic and viscous torques onto the inner core and determines its super-rotation \cite{GR96, ABO96}. Our present finding that the azimuthal rotation of the LSC has a sensitive dependence on its circulating strength may provide new insights as well as constraints for parameterization in advanced dynamo models. In future work there remain challenges to extend the intriguing problem of large-scale flow dynamics in rotating turbulent convection to broader parameter ranges relevant to geophysical and astrophysical fluid systems.          
   
This work was supported by the National Science Foundation of China through Grant No. 11572230 and 1561161004.


\begin{thebibliography}{29}
\bibitem[{com()}]{commentCESS}
\bibinfo{note}{In viewing that a general criteria of the amplitude threshold
  $\delta_c$ for cessations does not exist, we choose $\delta_c$=$0.5\delta_0$
  in order to ensure sufficient statistics to revel the LSC stochastic behavior
  under the influence of rotations. We note that the dynamical properties of
  cessations discussed in Fig. 4 can be well interpreted by the present model,
  independent on the threshold value of $\delta_c$ .}
\end{thebibliography}

\end{document}